\documentstyle[12pt,epsfig]{article}
\newcommand{\be}{\begin{equation}}
\newcommand{\ee}{\end{equation}}
\newcommand{\bea}{\begin{eqnarray}}
\newcommand{\eea}{\end{eqnarray}}
\newcommand{\lb}{\label}
\renewcommand{\theequation}{\arabic{section}.\arabic{equation}}

\def\cR{{\cal R}}

\def\cN{{\cal N}}

\def\e{{\rm e}}
\def\egg{{\rm e}^{-2\gamma\phi}}
\def\eg{{{\rm e}^{-\gamma\phi}}}

\def\s{{\sigma}}
\def\p{{\partial}}
\begin{document}
%\vskip-2cm
\title{
   \begin{flushright} \begin{small}
     LAPTH-806/00  \\ DTP-MSU/00-10 \\ hep-th/0007228
  \end{small} \end{flushright}
%\vspace{.5cm}
%%%%  Title %%%%
{\bf Solitons and black holes in Einstein-Born-Infeld-dilaton theory} }
%%%%%  Authors  %%%%
\author{
        G\'erard Cl\'ement
\thanks{Email: gclement@lapp.in2p3.fr}\\
{\it Laboratoire de  Physique Th\'eorique LAPTH (CNRS),}\\ {\it
B.P.110, F-74941 Annecy-le-Vieux cedex, France}\\
 and Dmitri Gal'tsov
\thanks{Email: galtsov@grg.phys.msu.su} \\
{\it Department of Theoretical Physics,}\\
        {\it Moscow State University, 119899, Moscow, Russia}
}
%%%%%  Address  %%%%
%\address
%{$^{a}$Laboratoire de  Physique Th\'eorique LAPTH (CNRS), \\
%B.P.110, F-74941 Annecy-le-Vieux cedex, France}
%\address
%{$^{b}$Department of Theoretical Physics,
%         Moscow State University, 119899, Moscow, Russia,}

%%%%%  Date  %%%%
\date{July 20, 2000}
\maketitle

\begin{abstract}
Static spherically symmetric asymptotically flat solutions to the $U(1)$
Born-Infeld theory coupled to gravity and to a dilaton are investigated.
The dilaton enters in such a way that the theory is $SL(2,R)$- symmetric
for a unit value of the dilaton coupling constant. We find globally regular
solutions for any value of the effective coupling which is the ratio of
the gravitational and dilaton couplings; they form a continuous
sequence labeled by the sole free parameter of the local solution
near the origin. The allowed values of this parameter are bounded from
below, and, as the limiting value is approached, the mass and
the dilaton charge rise indefinitely and tend to a strict equality
(in suitable units). Together with the electric charge they
saturate the BPS bound of the corresponding linear $U(1)$ theory.
Beyond this boundary the solutions become compact (singular),
while the limiting solution at the exact boundary value is
non-compact and non-asymptotically flat. The black holes in this
theory exist for any value of the event horizon radius and also
form a sequence labelled by a continuously varying parameter
restricted to a finite interval. The singularity inside the black
hole exhibits a power-law mass inflation.

\bigskip
Pacs no: 04.20.Jb, 04.50.+h, 46.70.Hg
\end{abstract}
%\begin{multicols}{2}
%\narrowtext

\newpage
%%%%%%%%%%%%%%%%%%%%%%%%%%%%%%%%%%%%%%%%%%%%%%%%%%%%%%%%%%%%%%%%%%%%%%
\section{Introduction}
%%%%%%%%%%%%%%%%%%%%%%%%%%%%%%%%%%%%%%%%%%%%%%%%%%%%%%%%%%%%%%%%%%%%%%
Born-Infeld (BI) type~\cite{BoIn34} generalisations of both
Abelian and non-Abelian gauge theories have recently attracted
much attention in the context of superstring theory. An Abelian BI
action was obtained as the disc open string effective action in an
external constant vector field~\cite{FrTs85,Ts86} (for references
and a recent review see \cite{Ts99}). The computation is valid
both for the bosonic string and the superstring and is exact in terms of
the sigma-model coupling $\alpha'$. Alternatively, the BI action
was derived as an effective action governing the dynamics of
vector fields on $D$-branes~\cite{Le89}. The BI lagrangian
introduces a natural bound for the field strength -- a Born-Infeld
`critical field' --  which should now be regarded as originating
 from the non-locality of the underlying fundamental theory. A
direct consequence of this is the smoothing of the pointlike
singularities of the vector field. As was shown already in the
original papers by Born and Infeld \cite{BoIn34}, the Coulomb
field of a point charge has a finite energy in this theory.

It is expected that a similar regularization of gravitational
singularities should follow from the non-locality of closed
strings~\cite{Ts94}. However no explicit closed string effective
action summing up all terms in $\alpha'$ was obtained so far. A
somewhat simpler question is how the singularity of the charged
black hole gets modified when gravity is still treated
classically, while the dynamics of the vector field  is governed
by the strings. The guess is that the singularity, although not
eliminated, will be smoothed somehow. As was shown by Gibbons and
Rasheed~\cite{GiRa95,Ra97}, the Einstein-Born-Infeld charged black
holes are less singular indeed as compared with their
Einstein-Maxwell counterpart, the Reissner-Nordstr\"om (RN)
solution. Namely, there is no RN-type divergent term in the metric
near the singularity $g_{00}\sim Q^2/r^2$, but only a
Schwarzschild type term $g_{00}\sim m_0/r$. But, contrary to the
original Schwarzschild solution, both signs of $m_0$ are possible
now for different values of the electric charge, thus both
timelike and spacelike singularities may appear. In the latter
case an internal Cauchy horizon is present as in the RN metric.

In fact, within the context of string theory two types of charged
black holes are known: besides the RN one there is also a
dilatonic black hole \cite{Gi82,GiMa88,GaHoSt91} with a rather
different internal structure. The extremal dilatonic black holes
are particularly different from the RN ones: the horizon is moved
to the singularity which is therefore lightlike. Moreover, in the
string frame the metric associated with the extremal magnetic
dilaton black hole turns out to be perfectly regular. Thus the
dilaton is also able to produce a regularizing effect on
gravitational singularities. A natural question arises concerning
the effect of both the BI non-linearity and the dilaton, this is
the main subject of the present paper. Since already the
flat space BI action gives rise to regular particle-like
solutions, we also expect such solutions in the
Born-Infeld-dilaton theory and its gravitating generalization. As
we will see, there is a one-parameter family of such solutions
which, for a limiting value of the parameter, tend (after suitable
coordinate rescaling) to the near horizon limit of the extremal
dilatonic black hole. The limiting solution saturates the BPS
bound. Another class of solutions is black holes. We show that a
generic Einstein-Born-Infeld-dilaton black hole has a timelike
singularity with a power-low behavior of the local mass function.

The choice of the lagrangian is worth to be discussed in detail.
We adopt the $SL(2,R)$ invariant version of the dilaton-axion
coupled Born-Infeld action \cite{GiRa96,Ra97,BrMoZu99}, which is a
direct Born-Infeld type generalisation of the toroidally compactified
heterotic string
effective theory ($N=4,\; D=4$ supergravity). The main reason is
that we would like to make contact with the dilatonic black
holes~\cite{Gi82,GaHoSt91}. This version of the theory is also
distinguished as the unique dilaton gereralization of the original
Born-Infeld theory exhibiting a continuous electric-magnetic
duality symmetry. In fact, the original BI theory as well as its
gravitating generalization is symmetric under the $SO(2)$
electric-magnetic duality rotations. This duality can be extended to the
$SL(2,R)$ symmetry when a dilaton and an axion are suitably added. However,
the open string version of the BI action coupled to a dilaton does
not enjoy electric-magnetic duality. This type of theory was
discussed along similar lines in a recent paper~\cite{TaTo00} with
results differing somewhat from those presented below. Meanwhile
the $SL(2,Z)\; S$-duality is now regarded as an exact symmetry of
the superstring/M theory, so it is reasonable to concentrate on
this version of the dilaton-coupled BI theory.

As was shown in~\cite{GiRa95,GiRa96}, the unique BI-dilaton-axion action
generating the $SL(2,R)$ invariant equations of motion reads (with
$G=\hbar=c=1$)
\bea \label{Sbiad}
S & = & \frac14\int \left\{- R +2(\nabla\phi)^2 + \frac12
\e^{4\phi}(\nabla\kappa)^2 - \kappa F_{\mu\nu}\tilde F^{\mu\nu}
+4\beta^2(1-\cR)\right\}\sqrt{-g}\,d^4x \nonumber \\ & = & \int L
\sqrt{-g} d^4x,
\eea
where  $\phi$ is a dilaton, $\kappa$ is an axion and
\be
\cR=\sqrt{1+\frac{F^2\e^{-2\phi}}{2\beta^2} - \frac{(F\tilde F)^2
\e^{-4\phi}}{16 \beta^4}},
\ee
with $F^2=F_{\mu\nu}F^{\mu\nu},\;F\tilde F={\tilde
F}_{\mu\nu}F^{\mu\nu}$. In the limit $\beta\to\infty$ this action
reduces to the (truncated) heterotic string effective action in
four dimensions.

In the BI theory, as in electrodynamics in media,
one must distinguish between the field tensor $F_{\mu\nu}$
incorporating the electric strength and the magnetic induction,
and the induction tensor $G_{\mu\nu}$ combining the electric
induction and the magnetic field strength
\be
G^{\mu\nu}=-\frac12 \frac{\p L}{\p
F_{\mu\nu}}=\frac{\e^{-2\phi}}{\cR} \left(F^{\mu\nu}-
\frac{(F\tilde F)}{4\beta^2}{\tilde F}^{\mu\nu} \e^{-2\phi}\right)
+\kappa {\tilde F}^{\mu\nu}.
\ee
The Maxwell equations in terms of $G_{\mu\nu}$ are sourceless
\be
dG=0, \quad G=G_{\mu\nu}dx^\mu\wedge dx^\nu,
\ee
thus coinciding with the Bianchi identity for the field tensor
\be
dF=0, \quad F=F_{\mu\nu}dx^\mu\wedge dx^\nu.
\ee
Therefore the linear transformations
\be
F\to a F+ b \tilde G,\quad G\to c G- d \tilde F,
\ee
do not change this system of equations. Moreover, one can check
that the equations for $\phi,\,\kappa$ following from the action
(\ref{Sbiad}) as well as the corresponding Einstein equations
\be
R_{\mu\nu}-\frac12 g_{\mu\nu}R=8\pi T_{\mu\nu},
\ee
\be
T_{\mu\nu}= G_{\mu\lambda}{F^{\lambda}}_\nu
+\phi_{,\mu}\phi_{,\nu}+\frac14
\kappa_{,\mu}\kappa_{,\nu}\e^{4\phi} -g_{\mu\nu}L
\ee
remain also unchanged~\cite{GiRa96}, provided the complex
axidilaton $z=\kappa+i \e^{-2\phi}$ undergoes the transformation
\be
z\to\frac{cz-d}{bz+a}
\ee
with real $a, b, c, d$ subject to $ac-bd=1$. Note that these
transformations change the action (\ref{Sbiad}) by a total
divergence.

In what follows we will consider either purely electric or purely
magnetic configurations. In this case the axion decouples and can
be consistently set to zero, with the term $F\tilde F$ in the
lagrangian omitted. This truncated version of the theory still has
a discrete electric-magnetic duality  $(b=-d=1, a=c=0)$:
\be \label{discrem}
F\to \tilde G,\quad G\to \tilde F,\quad \phi\to -\phi.
\ee
The plan of the paper is as follows. In Sec.~2 we perform the
dimensional reduction of the action for static spherically
symmetric purely electric (magnetic) configurations, and present
the system in two alternative gauges. Sec.~3 is devoted to
everywhere regular solutions which are studied both analytically
and numerically. Black hole solutions are discussed in Sec.~4. Our
results are summarized in Sec.~5. In the Appendix, the behavior of
the solutions near the singularities is discussed in more detail.
The numerical calculations presented in this paper were performed
in collaboration with V.V.~Dyadichev.

\setcounter{equation}{0}
%%%%%%%%%%%%%%%%%%%%%%%%%%%%%%%%%%%%%%%%%%%%%%%%%%%%%%%%%%%%%%%%%%%%%%
\section{Basic equations}
%%%%%%%%%%%%%%%%%%%%%%%%%%%%%%%%%%%%%%%%%%%%%%%%%%%%%%%%%%%%%%%%%%%%%%
We start with the action, which is a truncated (for purely
electric or magnetic configurations) version of (\ref{Sbiad})
\be \label{Sbid}
S=\frac14\int \left\{- \frac{R}{G} +2(\nabla\phi)^2 +
4\beta^2(1-\cR)\right\}\sqrt{-g}\,d^4x,
\ee
where we have restored the Newton constant $G$, and assume for
$\cR$ a more general form with an arbitrary dilaton coupling
constant $\gamma$:
\be
\cR=\sqrt{1+\frac{F^2\e^{-2\gamma\phi}}{2\beta^2}}.
\ee
The discrete electric-magnetic duality (\ref{discrem}) continues
to hold for an arbitrary $\gamma$. This theory has altogether
three dimensional parameters: $G$, $\beta$ and $\gamma$ with
dimensionalities (in units $\hbar=c=1$)
\be \label{dims}
[G]={\rm L}^2,\quad [\beta]={\rm L}^{-2},\quad [\gamma]={\rm L},
\ee
We will be dealing with static spherically symmetric
configurations, for which the equations of motion reduce to
one-dimensional equations. Assuming the general static spherically
symmetric parametrization of the metric
\be  \label{NsR}
ds^2=N\sigma^2 dt^2-\frac{dr^2}{N}-
R^2\left(d\theta^2+\sin^2\theta\, d\varphi^2\right),
\ee
where $N,\;R,\;\sigma$ are functions of $r$ only, we perform a
dimensional reduction leading to the Lagrangian
\be
L=L_g+L_m
\ee
where the gravitational part is
\be
L_g=\frac1G \left(N\sigma'RR'+\frac{\sigma}{2}((NR)'R'+1)\right),
\ee
while the matter part reads
\be
L_m=-\frac12 \sigma NR^2\phi'^2 +\sigma\beta^2 R^2(1-\cR).
\ee

The dual induction two-form is reduced to
\be
\tilde G=\tilde G_{\mu\nu}dx^\mu\wedge dx^\nu= \frac{a'R^2\egg
}{\sigma\cR}\sin\theta d\theta\wedge d\varphi
\ee
with
\be
\cR=\sqrt{1-\left(\frac{a'\eg }{\beta\s} \right)^2}.
\ee
Integrating the Maxwell equation $d\tilde G =0$ one obtains
\be
a'\egg=-\frac{Q V\s}{R^2},\quad \cR = V \equiv \left(1+
\frac{Q^2{\rm e}^{2\gamma\phi}}{\beta^2R^{4}}\right)^{-1/2}.
\ee

The dilaton equation reads
\be\label{dphi}
\left(\s NR^2\phi'\right)'=\s
R^2\gamma\beta^2\left(\frac1V-V\right).
\ee
Variation of the action over $R, \;N,\;\sigma$ gives the full set
of Einstein equations
\be\label{dR}
R(N\s' )'+\frac12 [\s(NR)']' +\frac12 N(\s R')'= G\left(2\beta^2\s
R (1-V)-\s NR\phi'^2 \right)
\ee
\be\label{dN}
R(\s R')' -2\s' RR' = - G\s R^2\phi'^2
\ee
\be\label{ds}
(NRR')'-\frac12 \left( R'(NR)' +1\right)=-G\left(\beta^2
R^2\frac{(1-V)}{V}+ \frac12
 NR^2\phi'^2\right).
\ee
Actually, the metric (\ref{NsR}) is invariant under
reparametrizations $r = r(\rho)$ of the radial coordinate, so that
a gauge condition may be imposed. We will use two particular
gauges in what follows: one is $R=r$, and another is $\s=1$.

In the first gauge ($\underline{R = r}$),
\be\lb{g1}
ds^2=N\sigma^2 dt^2-\frac{dr^2}{N}-
r^2\left(d\theta^2+\sin^2\theta\, d\varphi^2\right)\,.
\ee
Introducing a new variable $\psi = \beta^2\gamma^4 Q^2 {\rm
e}^{2\gamma\phi}$, and performing rescalings
\be\lb{rescal}
\beta\gamma r\to r,\quad \gamma \phi \to \phi,
\ee
(leading to a dimensionless independent variable $r$), the system
of equations (\ref{dphi}), (\ref{dN}), (\ref{ds}) may be rewritten
as
\bea
Nr(r\phi')' +r\phi' &=&\frac{\psi}{\sqrt{\psi+r^4}}+ 2g
r\phi'\left( \sqrt{\psi+r^4} - r^2 \right) \lb{phir}\\
(rN)'-1&=&g\left( 2r^2-2\sqrt{\psi+r^4} -Nr^2\phi'^2 \right)
\lb{Nr} \\ \s'&=&g\s r\phi'^2\,, \lb{sr}
\eea
with the only dimensionless coupling constant
\be
g=\frac{G}{\gamma^2}\,.
\ee
Another useful relation is the combination $2R$(\ref{dR}) $-
2N$(\ref{dN}) which gives, after rescaling,
\be\lb{R00}
\left(\frac{(N\s^2)'r^2}{\s}\right)' = 4g\s r^2(1-V)\,.
\ee

In the second gauge ($\underline{\s = 1}$),
\be\lb{g2}
ds^2 = \lambda^2 dt^2-\frac{d\rho^2}{\lambda^2}-
R^2\left(d\theta^2+\sin^2\theta\, d\varphi^2\right)\,,
\ee
the radial coordinate $\rho$ being related to the radial
coordinate $r$ in the first gauge by $|d\rho/dr| = \sigma(r)$. We
note that a suitable combination of Eqs. (\ref{dphi}), (\ref{dR})
and (\ref{ds}) eliminates the source terms, leading to the
equation
\be
2(\lambda^2 RR')' - \frac{1}{2}(\lambda^2R^2)'' +
\frac{2G}{\gamma}(\lambda^2R^2\phi')' - 1 = 0\,.
\ee
The first integral of this equation, together with equations
(\ref{dR}) and (\ref{dphi}), leads (after rescaling $\rho$ and
$R$, as well as $\phi$, as before), to the system
\bea
\frac{R'}{R} - \frac{\lambda'}{\lambda} + 2g\phi' & = & \frac{\rho
- \rho_0} {\lambda^2R^2}\,, \lb{first}
\\
R''& = & - gR\phi'^2\,, \lb{Rrho}\\ (\lambda^2R^2\phi')' & = & u
\;\equiv V\frac{\psi}{R^2}\,, \lb{phirho}
\eea
with $\rho_0$ an integration constant.

\setcounter{equation}{0}
%%%%%%%%%%%%%%%%%%%%%%%%%%%%%%%%%%%%%%%%%%%%%%%%%%%%%%%%%%%%%%%%%%%%%%
\section{Regular solutions}
%%%%%%%%%%%%%%%%%%%%%%%%%%%%%%%%%%%%%%%%%%%%%%%%%%%%%%%%%%%%%%%%%%%%%%

In the gauge $R=r$ the conditions of asymptotic flatness read, in
terms of rescaled quantities,
\bea\lb{as}
N&=&1-\frac{2M}{r}+O(\frac{1}{r^2})\,,
\\ \phi&=& \phi_\infty
+\frac{D}{r}+O(\frac{1}{r^2})\,,
\\ \sigma&=&
1-\frac{gD^2}{2r^2}+O(\frac{1}{r^3})\,.
\eea
Here $M$ and $D$ are the  mass and the dilaton charge in the
rescaled variables. Alternatively, one can introduce the unscaled
quantities $\overline{M}$ (via $N = 1 - 2G\overline{M}/r+ ...$)
and $\overline{D}$, then $M =G \beta\gamma\overline{M}$, $D =
\beta\overline{D}$.

For these quantities one can derive useful sum rules. Define a
local mass function $m(r)$ through
\be
N=1-\frac{2m(r)}{r},
\ee
with $m(r) \to M + O(1/r)$ as $r \to\infty$. Integrating Eq.
(\ref{R00})
 from some finite $r_1$ to infinity and taking into account the
conditions of asymptotic flatness one obtains
\be  \lb{Mr1}
M = 2g\int_{r_1}^\infty \s r^2(1-V) dr+\frac{(N\s^2
)'r^2}{2\s}\Big |_{r=r_1}\,.
\ee
For solutions regular at the origin the last term vanishes as
$r_1\to 0$, so one obtains
\be \label{M}
M = 2g\int_{0}^\infty \s r^2(1-V) dr\,.
\ee
In a similar way one can derive from the dilaton equation
\be \label{Dr1}
D = \int_{r1}^\infty \s r^2\left(V-V^{-1}\right)dr -\sigma N r
\xi\Big|_{r=r_1}\,,
\ee
with $\xi = r\phi'$. Thus for the regular solutions we have
\be \label{D}
D = \int_{0}^\infty \s r^2\left(V-V^{-1}\right)dr\,.
\ee
Combining these two formulas we find the relation
\be \label{MD}
M + gD = - g\int_{0}^\infty \s r^2\frac{(1-V)^2}{V}dr\,,
\ee
showing that the dilaton charge for the electric solution is
negative ($V < 1$) and that its absolute value is greater than the
mass.

To analyse the behavior of regular solutions near the origin it is
convenient to introduce the logarithmic variable $\tau=\ln r$. The
system (\ref{phir})--(\ref{sr}) may be rewritten as the system of
first order equations
\bea
\dot\psi&=&2\xi\psi\,, \lb{psi} \\ N\dot\xi&=&u-\xi v\,,  \lb{xi}
\\ \dot N&=&v-N(1+g\xi^2)\,, \lb{N} \\ \dot\s&=&g\s\xi^2\,,
\lb{sig}
\eea
where
\be
u=\frac{\psi}{\sqrt{\psi+r^4}},\quad
v=1+2g\left(r^2-\sqrt{\psi+r^4}\right)\,.
\ee

By definition, the regular solution should be flat in the vicinity
of the origin $r = 0$ ($\tau \to -\infty$), i.e. $N=1$ at the
origin. Under this condition one can find the series solution in
powers of $1/\tau$ with logarithmic coefficients
\bea\lb{reg}
\psi&=&\frac{1}{\tau^2}\left(1+\frac{2L}{\tau}+
\frac{3L^2-2(2g-1)L+4(1-g)}{\tau^2}+\dots \right)\,,
\label{erpsi}\\ \xi&=&\frac{1}{\tau}\left(-1+\frac{-L + 2g
-1}{\tau}- \frac{L^2-3(2g-1)L+4(1-g)^2+1}{\tau^2}+\dots\right)\,,
\label{erxi}\\ N&=&1+\frac{2g}{\tau}\left(1 + \frac{L+1/2}{\tau} +
\frac{L^2+2(1-g)(L+3/2)}{\tau^2}+\dots\right)\,. \label{erN}
\eea
In these formulas
\be
L=(2g-1)\ln (-\tau) +c,
\ee
where $c$ is a free parameter. One can also show from Eq.
(\ref{R00}) that $(N\sigma^2)\dot{}\simeq (4g/3) \sigma e^{2\tau}$
is vanishingly small, so that $\lambda^2 = N\sigma^2$ is constant
to all orders in this expansion. Two particular values of $g$
should be mentioned. The first one is $g=0$ which corresponds to a
vanishing Newton constant. In this case the expansion (\ref{erN})
generates $N\equiv 1$, so we get the flat space dilaton-BI
solution. The second is $g=1/2$, in this case there are no
logarithmic terms in the expansions.

These local solutions being continued (numerically) to large $r$
meet the conditions of asymptotic flatness for continuously
varying $c$ subject to the condition
\be
c>c_{cr}(g)\,.
\ee
The dependence of $c_{cr}$ on the effective coupling constant $g$
is shown in Fig.1. The limit $g=0$ corresponds to the flat space
solutions, in this case $N\equiv 1$. The corresponding value is
approximately $c_{cr}=1.52$. When $g$ is large enough the critical
value moves to the negative half-plane. The variation with
$c_{cr}$ of the mass, dilaton charge and the asymptotic value of
$\psi$ are shown in Fig.~2. It is worth emphasizing that we deal
with a one-parameter family, i.e. for a given mass the values of
the dilaton charge and the electric charge (proportional to the
square root of $\psi_{\infty}$) are fixed. Physically these
particle-like solutions can be regarded as the original
Born-Infeld field distributions of the point charge deformed by
gravity and by the dilaton. Thus the mere existence of regular
particle-like solutions in the present theory is by no means
surprising.

When the parameter $c$ decreases and approaches the critical
value, both the mass and the absolute value of the dilaton charge
increase, with an approximate power law dependence $M \propto
(c-c_{cr})^p$, where $ p = 1/(1+g)$ (see Fig.~2), and at the same
time
\be\lb{BPS}
g|D|\to M \quad {\rm as} \quad c\to c_{cr}\,.
\ee
This can be understood qualitatively using the sum rules above.
When $c$ is close to its boundary value the main contribution to
the integral comes
 from the region of large $r$ where $V\sim 1$. Then from Eqs.
(\ref{M}), (\ref{D}), (\ref{MD}) it is clear that if the integral
for $M$ diverges for $r \to \infty$, the integral for the sum
$M+gD$ is less divergent than both $M$ and $|D|$. The numerical
results also show that the ratio $\psi(\infty)/MD$ goes to a
finite limit as $c \to c_{cr}$. To understand this, note that a
generic asymptotically flat solution of the non linear theory is
asymptotic for $r \to \infty$ ($V \to 1$) to an electric solution
of the linear Einstein-Maxwell-dilaton theory (obtained
 from the magnetic solution of \cite{GaHoSt91} by transforming $\phi \to -\phi$
and rescaling $\rho$, $R$ and $\phi$)
\bea\lb{GHS}
\psi & = & \frac{\rho_+\rho_-}{g+1} \left(1 - \frac{\rho_-}{\rho}
\right)^{2/(g+1)}\,, \quad R = \rho\left(1 -
\frac{\rho_-}{\rho}\right)^{1/(g+1)}\,, \nonumber \\ \lambda^2 & =
& \left(1 - \frac{\rho_+}{\rho}\right) \left(1 -
\frac{\rho_-}{\rho}\right)^{(g-1)/(g+1)}
\eea
($g = \gamma^{-2}$ for $G = 1$). Comparing with (\ref{as}), we see
that $M + gD = (\rho_+ - \rho_-)/2$, so that the limit $M + gD \ll
M$ corresponds to the BPS limit $\rho_- \to \rho_+$. In this
limit, we obtain from (\ref{GHS})
\be \lb{psinfty}
\psi(\infty) \simeq (g+1)\left(\frac{M}{g}\right)^2\,.
\ee
This relation is well confirmed by the numerical observations
(Fig. 2). Relations (\ref{psinfty}) and (\ref{BPS}) together imply
\be
M^2+g^2 D^2=Q_{eff}^2,\quad Q_{eff} =
g\left(\frac{2}{g+1}\,\psi(\infty)\right)^{1/2}\,,
\ee
which corresponds to the BPS saturation of the
Einstein-Maxwell-dilaton theory~\cite{Gi82}.

The transition point in the parameter space $c=c_{cr}$ corresponds
to a limiting solution which is not asymptotically flat. To find
its asymptotic behaviour, consider the asymptotic solution
(\ref{GHS}) in the BPS limit $\rho_- \to \rho_+$. The limit in
which both $\rho$ and $M = g\rho_+/(g+1)$ go to infinity
corresponds to the near horizon limit $|\rho - \rho_+| \ll
\rho_+$. Putting $ (\rho - \rho_+)/\rho_+ = (r/\rho_+)^{g+1}$,
rescaling $t$, and taking the limit $\rho_+ \to\infty$, we arrive
at the asymptotic critical solution
\begin{equation}\label{cr}
ds_{cr}^2 = a^2 r^{2g}dt^2 - (g+1)^2 dr^2 - r^2d\Omega^2\,, \quad
\psi_{cr} = r^2/(g+1)\,.
\end{equation}
This asymptotic behaviour has not been directly tested, as in all
our numerical computations the parameter $c$ was at best only
approximately equal to its critical value. Nevertheless the
critical solution can be approached numerically as the envelope of
asymptotically flat regular solutions with $c \simeq c_{cr}$. In
this respect, we note that the curves for $\xi(\tau)$ show
increasingly large and flat maxima near $\xi = 1$ (before dropping
again to 0 as $\tau \to \infty$), in agreement with the asymptotic
critical behaviour $\xi_{cr}(\infty) = 1$ from (\ref{cr}).

For $c<c_{cr}$ the solution starting at the origin as
(\ref{erpsi},\ref{erxi},\ref{erN}) develops a coordinate
singularity at some radius $r=r^*$ . In the gauge $R=r$ one
observes that for $r \to r^*$ the metric function $N$ approaches
zero while the dilaton has a square root singularity
\be
N\sim (r^*-r)\,,\quad \phi\sim \sqrt{r^*-r}\,, \quad \xi\sim
\frac{1}{\sqrt{r^*-r}}.
\ee
Such a behaviour can be derived by defining locally a new radial
coordinate $x$:
\be\lb{x}
r = r^* \exp(- gx^2/2)\,,
\ee
and assuming $\psi \simeq \psi^* (1 - 2bx)$, where $b$ is some
constant. Then $\xi \simeq b/gx$, and Eq. (\ref{xi}) yields $N
\simeq - gv^* x^2$, with $v^* < 0$ ($g\psi^* > r^{*2} + 1/4g$) by
continuity, while Eq. (\ref{N}) yields $b^2 = 1$, so that $b = \pm
1$. The complete local solution
\bea\lb{bg}
\xi & = & \pm(1/gx) + d + ... \,, \nonumber \\ N & = & gx^2(-v^*
\mp g(u^* + dv^*)x + ...) \,, \\ \s & = & k(1/x \mp 2gd + ...)
\nonumber
\eea
depends on the four integration constants $r^*$, $\psi^*$ $d$ and
$k$, and so is generic.

Obviously the  coordinate patch $r < r^*$ fails to describe the
full solution. The extension may be achieved by going to the
second gauge (\ref{g2}), as from (\ref{x}) and the last equation
(\ref{bg}) $d\rho \propto dx$ near $R \equiv r = r^*$. From Eq.
(\ref{phirho}), $R^2\lambda^2\phi'$ is an increasing function of
$\rho$. So if one starts from a solution regular near $R = 0$
($\phi'_{\rho} \propto \phi'_r > 0$) and increases $R$ and $\rho$,
one must have $\phi'_{\rho} > 0$ at $R = r^*$, with $R'_{\rho} =
0$. This is a maximum of $R$ from Eq. (\ref{Rrho}). As $\rho$ is
further increased, $R$ must decrease to zero, while $\phi'_{\rho}$
stays positive so that $\phi$ continues to increase, and a regular
solution at $R = 0$ ($\phi \to - \infty$) cannot be achieved. The
resulting solution is compact and singular (`bag of gold'), with
the power law behaviours near the singularity
\bea
\psi \propto r^{2\delta}\,,& \quad N \propto r^{-1 -g\delta^2}\,,&
\quad \s \propto r^{g\delta^2} \qquad \;(g > 1/4)\,, \lb{sng1} \\
\psi \propto r^{-1/g}\,,& \quad N \propto r^{-1/2g}\,,\; & \quad
\s \propto r^{1/4g} \qquad (g < 1/4)\,, \lb{sng2}
\eea
where $\delta < 0$ is bounded by
\be\lb{Bg}
\sqrt{4 + 3g^{-1}} - 2 < -\delta < \sqrt{g^{-1}}\,.
\ee
(details on the derivation of these behaviours and bounds are
given in the Appendix).

Let us consider now magnetically charged solitons. Reverting to
unscaled variables and introducing a magnetic charge via
\be
F=P\sin\theta d\theta\wedge d\varphi
\ee
we obtain
\be
\cR=\sqrt{1+\frac{P^2\egg}{\beta^2 R^2}}
\ee
Rescaling the variables in the gauge $R=r$ we obtain the same
system (\ref{phir})--(\ref{sr}) with $\xi\to -\xi,\; Q\to P$. This
is in a perfect agreement with the electric-magnetic duality
discussed in the Introduction. Accordingly, the electric solitons
described in this section can be reinterpreted as magnetic
solitons. In this case the parameter  $\delta $ in (\ref{sng1}) is
positive, and it is easy to see that the bounds above still hold
if $-\delta$ is replaced by $|\delta|$. It is worth noting that
the above symmetry holds in the Einstein frame, in the conformally
related `string' frame the metrics of electric and magnetic
solutions are essentially different.

\setcounter{equation}{0}
%%%%%%%%%%%%%%%%%%%%%%%%%%%%%%%%%%%%%%%%%%%%%%%%%%%%%%%%%%%%%%%%%%%%%%
\section{Black holes}
%%%%%%%%%%%%%%%%%%%%%%%%%%%%%%%%%%%%%%%%%%%%%%%%%%%%%%%%%%%%%%%%%%%%%%

To study black holes, let us assume the existence of a
non-degenerate horizon at some $r = r_h$, i.e. $N(r_h) = 0$, with
$N'(r_h) > 0$. It is again convenient to work with the rescaled
first order differential system (\ref{psi}-\ref{sig}) with the
logarithmic variable $\tau = \ln(r/r_h)$. From this system we find
the following series solution
\bea
N&=&v_h \tau+\frac{1}{2v_h}(v_1v_h-v_h^2-gu_h^2)\tau^2+O(\tau^3),
\lb{eNh}  \\ \xi &=&   \frac{u_h}{v_h} +\frac{1}{2v_h^2}(v_h
u_1-v_1 u_h)\tau+ O(\tau^2), \lb{exih}  \\ \psi &=&
b+\frac{2bu_h}{v_h}\tau +O(\tau^2),\lb{epsih}
\eea
Here $u_h=u(r_h)$,
\be
v_1=2g\left(2r_h^2-\frac{bu_h+2v_h
r_h^4}{v_h\sqrt{b+r_h^4}}\right),\quad
u_1=\frac{b}{(b+r_h^4)^{3/2}}\left(2r_h^4\left(\frac{u_h}{v_h}-1\right)
+ b\frac{u_h}{v_h}\right)\,,
\ee
and $b=\psi(r_h)$ is a free parameter varying in the finite
interval
\be
-r_h^4<b<\frac1g\left(\frac{1}{4g} +r_h^2\right)\,.
\ee
The left bound corresponds to the condition of positivity under
the square root in $V$, while the right bound comes from the
assumption $v_h > 0$.

Actually the right bound on the parameter $b$ for asymptotically
flat solutions is lower, namely
\be
b<b_{cr}(g)< \frac1g\left(\frac{1}{4g} +r_h^2\right)\,,
\ee
otherwise the bag of gold type singularity is met. The critical
$b$ depends on the horizon radius as well. On Fig.~4 the curves
$b_{cr}(g)$ are shown for various values of $r_h$.  For small
$r_h$ the exterior black hole solutions approach regular
solutions. As in the case of regular solutions with $c$ close to
$c_{cr}$, both the mass and the dilaton charge grow with $b$. From
the Eqs.~(\ref{Mr1}) and (\ref{Dr1}) one can see that the boundary
term in the black hole case vanishes for the dilaton but remains
finite for the mass:
\be  \lb{Mrh}
M =M_0+ 2g\int_{r_h}^\infty \s r^2(1-V) dr,\quad M_0=\frac{v_h
\s_h r_h^2}{2}
\ee
Here the first term may be regarded as the 'bare' mass of the
black hole, and the second as a contribution from the black hole
hair. The sum rule for the dilaton charge preserves its form
(\ref{D}), where the integration now is performed over the
exterior space:
\be\lb{Dh}
D = \int_{r_h}^\infty \s r^2\left(V-V^{-1}\right)dr
\ee
The combined sum rule now reads
\be \label{MDh}
M-M_0 + gD = - g\int_{r_h}^\infty \s r^2\frac{(1-V)^2}{V}dr.
\ee
When $b_{cr}$ is approached this quantity remains finite while $M$
and $D$ diverge, so that asymptotically $g|D|$ tends to the field
mass $M-M_0$.

 From Eq. (\ref{Rrho}), $R(\rho)$ must vanish for some $\rho_s
< \rho_h$, leading to a singularity. There can be no inner Cauchy
horizon, so that this singularity must be timelike. To show this,
note that from Eq. (\ref{phirho}), $\lambda^2R^2\phi'$ must
decrease when $\rho$ decreases inside the event horizon $\rho =
\rho_h$, and thus must stay negative ($\phi_h'$ is positive, while
$\lambda^2 < 0$ for $r < r_h$), so that $\lambda^2$ cannot vanish
again. Reversing the sign of $\tau$ in the
Eqs.~(\ref{psi})--(\ref{sig}) one can integrate inside the black
hole up to the singularity. The leading terms near the singularity
are the same as in Eq. (\ref{sng1}), with now $\delta$ positive
for electric black holes and negative for magnetic black holes,
\be \label{es}
m \simeq \frac{\mu}{r^{g\delta^2}},\quad \xi \simeq \delta\,,
\quad \psi \simeq \nu^2 r^{2\delta}\,, \quad \s \simeq
ar^{g\delta^2}
\ee
where $\delta$, $\mu > 0,\; \nu > 0$ and $a > 0$ are free
parameters. The solution (\ref{es}) is generic, so starting from
the expansions (\ref{eNh})--(\ref{epsih}) inside the horizon one
always meets a member of the local family (\ref{es}) with certain
$\mu,\nu, a, \delta$.

The parameter $\delta$ is constrained by various bounds (see the
Appendix). One of these is obtained from the combined sum rule
similar to (\ref{MDh}), but with the integration covering the
internal space,
\be\lb{MDs}
\mu a (1 - g\delta^2 + 2g|\delta|) - M_0 = g \int_0^{\rho_h} R^2
\frac{(1 - V)^2}{V} d\rho \,.
\ee
On account of the positivity of $M_0$ and $\mu a$, this implies $1
- g\delta^2 + 2g|\delta| > 0$, i.e.
\be\lb{Bh1}
|\delta| < 1 + \sqrt{1+g^{-1}} \,.
\ee
Another bound, obtained by combining the sum rule (\ref{MDs}) with
the first integral (\ref{first}) evaluated at the singularity, is
\be\lb{Bh2}
g|\delta| < \rho_h/2\mu a - 1\,.
\ee
In the limit $g \to \infty$, corresponding to $\gamma \to 0$, the
dilaton decouples, and the bound (\ref{Bh2}) leads, for fixed
($\rho_h$, $\mu$, $a$) (note that the ratio $\rho_h/\mu a$ is
invariant under the rescaling (\ref{rescal})), to $\delta =
O(g^{-1})$, so that the exponents in (\ref{es}) go to zero, and
the mass function goes to a constant at the singularity $r = 0$,
consistently with the results of \cite{GiRa95,Ra97} for
Einstein-Born-Infeld black holes \footnote{Note that (\ref{MDs})
implies $\mu > 0$ in the limit $g \to \infty$. Here there is only
one horizon for finite $g$, so the two-horizon case ($\mu < 0$)
cannot be obtained as a limit.}. Thus the dilaton reinforces a
divergence of the mass function at the singularity.

\setcounter{equation}{0}
%%%%%%%%%%%%%%%%%%%%%%%%%%%%%%%%%%%%%%%%%%%%%%%%%%%%%%%%%%%%%%%%%%%%%%
\section{Conclusion}
Let us summarize our results. We have shown that the
Einstein-Born-Infeld-dilaton theory admits a one-parameter family
of particle-like globally regular solutions characterized by their
mass, electric (magnetic) charge, and dilaton charge. The unique
parameter determining these three quantities is bounded from
below, and when the boundary is approached, the charges exhibit a
BPS saturation similar to that of the BPS black holes of
Einstein-Maxwell-dilaton theory (recall that the latter does not possess
particle-like solutions). In this limit the absolute values of
charges rise indefinitely, as well as the effective radius of the
particle, so that the main contribution to the total charges comes
 from large radii. Since at large radii the Born-Infeld
non-linearity is small, the correspondence with the Maxwellian
counterpart of the theory is by no means surprising. However the
BPS saturation property of regular solutions at large masses is not obvious 
{\it a priori}.

Another type of solutions regular at the origin are compact, with
a point singularity at the `other end' (such configurations are
commonly called `bags of gold
c'). For these solutions, the areas of
two-sphere sections increase up to some finite limiting value and
then decrease again up to the final curvature singularity. These
solutions also form a one-parameter sequence, corresponding to
values of the parameter below the threshold for particle-like
solutions.

We have also found a two-parameter family of black-hole solutions for
arbitrary values of the horizon radius and the second free parameter varying
on a finite interval. Alternatively, one can consider as independent
parameters the mass and the electric (magnetic) charge of the black hole. 
The dilaton charge
is a dependent quantity like in the linear Einstein-Maxwell-dilaton theory.
Altogether these three quantities satify a BPS inequality which is saturated
in the limit of the infinite mass and the dilaton charge. 
The black holes do not possess internal Cauchy horizons and the
singularity inside them is always spacelike. This has to be 
contrasted with the Born-Infeld black holes without dilaton in which case
the solutions possessing  an internal Cauchy horizon still exist. 
Like in the linear Maxwell-dilaton case, the scalar field prevents the 
formation of an internal horizon. The local mass function has a power-low 
behaviour near the singularity with a power index depending on the global 
black hole parameters. Solutions with a regular
event horizon exist also in the bag of gold (compact) form, these
are doubly singular.

The above picture holds in the Einstein frame. One should inquire
how the regular solutions look in the 'string' frame which is related
to the present one by a conformal transformation with the conformal factor
$\psi$. According to (\ref{erpsi}), this conformal factor vanishes 
logarithmically as $r\to 0$ for electric solutions (and diverges for magnetic 
ones), so the 'string' metric in the vicinity of the origin is 
singular. Nevertheless, the ADM mass remains finite in this frame both 
for electric and magnetic solutions.

It is worth comparing our results with those of the paper
\cite{TaTo00} where another type of coupling of the Born-Infeld
theory to dilaton was assumed. In the latter case there is no
symmetry between electric and magnetic solutions, in particular,
the magnetic non-singular particle-like solutions were found 
in the string frame (contrary to electric ones). Our version
of the theory is (classically) $S$-dual and, moreover, it inherits
the structure of the $\cN=4,D=4$ supergravity. Therefore 
in the Einstein frame magnetic solutions are related to 
electric ones simply by reversing the sign of the dilaton.
The string metric is singular at the origin in both cases.

\bigskip
\noindent {\large {\bf Acknowledgements}}

\bigskip
We would like to thank V.V.~Dyadichev for assistance in numerical
calculations. D.G. is grateful to Laboratoire de Gravitation et 
Cosmologie Relativiste, Universit\`e  Paris-6 
for hospitality and to CNRS for support while the work was initiated.
His work was also partially supported by the Russian Foundation for
Basic Research under the grant 00-02-16306.
%%%%%%%%%%%%%%%%%%%%%%%%%%%%%%%%%%%%%%%%%%%%%%%%%%%%%%%%%%%%%%%%%%%%%%

\section*{Appendix: Power law behaviours near the singularity}
\appendix
\def\theequation{A.\arabic{equation}}
\setcounter{equation}{0}

Let us first consider the electric bag of gold of Sect. 3 with a
regular origin and a point singularity. Assume the power law
behaviours
\be \lb{sng0}
\psi \simeq \nu^2 r^{2\delta}\,, \quad \s \simeq a r^{g\delta^2}
\ee
(with the integration constants $\delta < 0$, $\nu > 0$, $a > 0$).
Then, if
\be\lb{Bg0}
g\delta^2 + \delta + 1 > 0\,,
\ee
Eqs. (\ref{phir})-(\ref{Nr}) are solved near the singularity $r =
0$ by
\be\lb{sngN1}
N \simeq -2\mu r^{-1-g\delta^2}\,,
\ee
with $\mu < 0$. Several bounds on the exponent $\delta$ can be
derived. First, Eq. (\ref{first}) where $\rho_0 = -(\lambda^2
R/\s)(1+2g\xi)|_{r=0} = 0$ for regular solutions gives, near the
singularity $\rho = \rho_1$,
\be\lb{Bg1}
g\delta^2 - 4g\delta - 3 = -\rho_1/\mu a > 0\,,
\ee
leading to the bound
\be\lb{Bg2}
- \delta > \sqrt{4 + 3g^{-1}} - 2\,.
\ee
Other bounds can be obtained from sum rules obtained in a manner
similar to (\ref{M}), (\ref{D}), (\ref{MD}). These sum rules are
\bea
\mu a (g\delta^2 - 1) & = & 2g \int_0^{\rho_1} R^2 (1 - V) d\rho
\,, \lb{Mg}\\ 2\mu a \delta & = &  \int_0^{\rho_1} R^2 (V^{-1} -
V) d\rho \,, \lb{Dg}\\ \mu a (1 - g\delta^2 + 2g\delta) & = & g
\int_0^{\rho_1} R^2 \frac{(1 - V)^2}{V} d\rho \,, \lb{MDg}
\eea
where all the right-hand sides are finite and positive. From Eqs.
(\ref{Mg}) and  (\ref{MDg}) we obtain the bounds
\be\lb{Bg3}
\sqrt{1 + g^{-1}} - 1 < -\delta < \sqrt{g^{-1}}\,,
\ee
while Eq. (\ref{Dg}) gives nothing new. It is easy to show that
the lower bound (\ref{Bg3}) is weaker than the bound (\ref{Bg2}),
so that our final bounds are
\be\lb{Bg4}
\sqrt{4 + 3g^{-1}} - 2 < -\delta < \sqrt{g^{-1}}\,.
\ee
A corollary of Eq. (\ref{Bg4}) is that the singular behaviour
(\ref{sng1}) can exist only for
\be
g > 1/4\,,
\ee
which ensures the inequality (\ref{Bg0}). Also, the combination of
(\ref{Bg1}) and (\ref{MDg}) leads to the inequality
\be
-g\delta < 1 - \rho_1/2\mu a \,.
\ee

Assume now the power law behaviours (\ref{sng0}) with
\be
g\delta^2 + \delta + 1 < 0\,,
\ee
which is possible only for
\be
g < 1/4\,.
\ee
Then the behaviour
\be\lb{sngN2}
N \simeq \alpha r^{\delta}
\ee
solves Eq. (\ref{Nr}) if $\alpha = -2g\nu/(1+\delta+g\delta^2) >
0$, and Eq. (\ref{phir}) if $1 + 2g\delta= 0$, leading to the
behaviours (\ref{sng2}). Eq. (\ref{first}) does not give new
information, neither do the sum rules analogous to (\ref{Dg}) and
(\ref{MDg}) whose both sides diverge with the same power law and
sign, while the sum rule analogous to (\ref{Mg}) shows that the
behaviour of $\lambda^2$ near the singularity must be $\lambda^2
\simeq \alpha a^2 - \beta(\rho_1 - \rho)^{\gamma}$, with $\gamma =
(1-4g)/(1+4g) > 0$, $\beta > 0$. For $g = 1/4$, (\ref{sngN2}) is
replaced by $N \simeq - (\nu/2)\,r^{-2}\ln r$.

Consider now the black hole case. Let us assume again the
behaviours (\ref{sng0}), with $\delta > 0$ for electric black
holes. In this case $g\delta^2 + |\delta| + 1 > 0$ for all $g \ge
0$, so that the behaviour of $N$ is always given by (\ref{sngN1})
with $\mu > 0$. As in the case of the bag of gold singularity, we
can derive various bounds on $|\delta|$. On the horizon Eq.
(\ref{first}) gives
\be
\rho_0 - \rho_h = r_h^2(\lambda^2)'(\rho_h)/2 = M_0
> 0 \,.
\ee
Applying now the same equation (\ref{first}) near the singularity
$\rho \propto r^{g\delta^2+1} \to 0$ and inserting the preceding
result, we obtain to leading order
\be\lb{Bh3}
g\delta^2 - 4g\delta - 3 = - \frac{(\rho_h + M_0)}{\mu a}\,.
\ee
The negativity of the right-hand side leads to the bound
\be\lb{Bh4}
\delta < 2 + \sqrt{4 + 3g^{-1}}\,.
\ee
On the other hand, the sum rules obtained as in the bag of gold
case give
\bea
\mu a (g\delta^2 - 1) + M_0 & = & 2g \int_0^{\rho_h} R^2 (1 - V)
d\rho \,, \lb{Mhs}\\ 2\mu a \delta & = &  \int_0^{\rho_h} R^2
(V^{-1} - V) d\rho \,, \lb{Dhs}\\ \mu a (1 - g\delta^2 + 2g\delta)
- M_0 & = & g \int_0^{\rho_h} R^2 \frac{(1 - V)^2}{V} d\rho \,.
\lb{MDhs}
\eea
This last sum rule (with $M_0$ and $\mu a$ positive) leads to the
bound
\be\lb{Bh5}
\delta < 1 + \sqrt{1 + g^{-1}}\,,
\ee
which is stronger than the bound (\ref{Bh4}). Again, the
combination of (\ref{Bh3}) and (\ref{MDhs}) leads to the
inequality
\be
g\delta < \rho_h/2\mu a - 1\,.
\ee

%%%%%%%%%%%%%%%%%%%%%%%%%%%%%%%%%%%%%%%%%%%%%%%%%%%%%%%%%%%%%%%%%%%%%%%%%%%%

%\end{multicols}
\newpage

%%%%%%%%%%%%%%%%% c_cr(g) for reg
\begin{figure}
\unitlength1cm
\begin{picture}(14,11)
\put(1,0){\epsfig{file=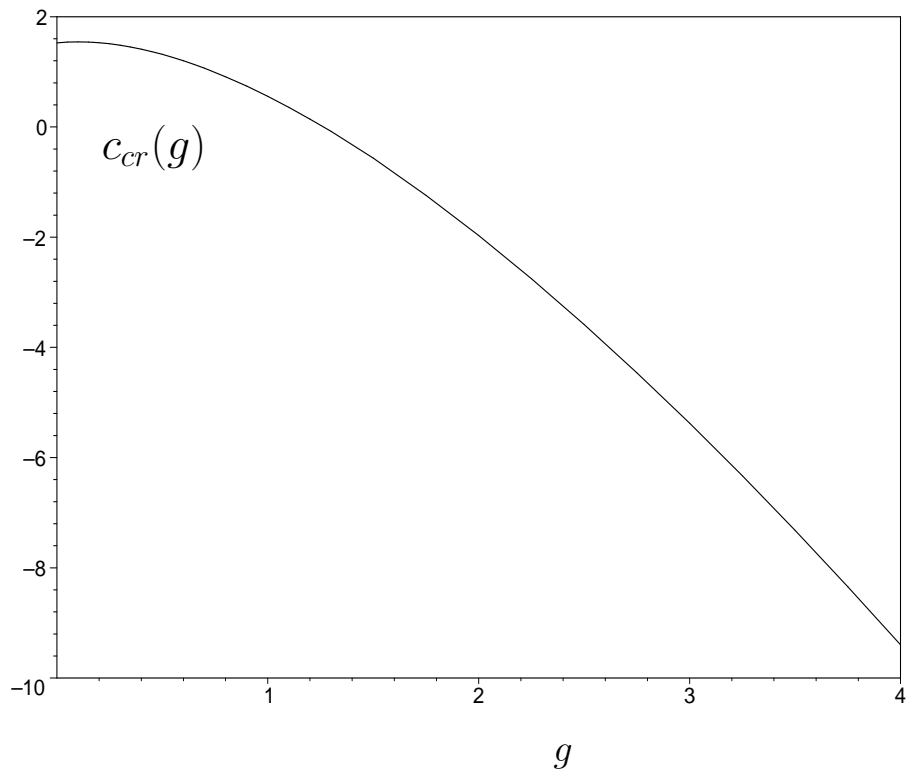,width=14cm,height=11cm}}
\put(3,8){{\Large $c_{cr}(g)$}} \put(9,0){{\large $g$}}
\end{picture}

\bigskip
\caption{Critical parameter $c_{cr}$ as a function of the
dimensionless coupling constant $g$. } \label{c_cr}

\end{figure}

%%%%%%%%%%%%%%%%% BPS limit
\begin{figure}
\unitlength1cm
\begin{picture}(14,11)
\put(1,0){\epsfig{file=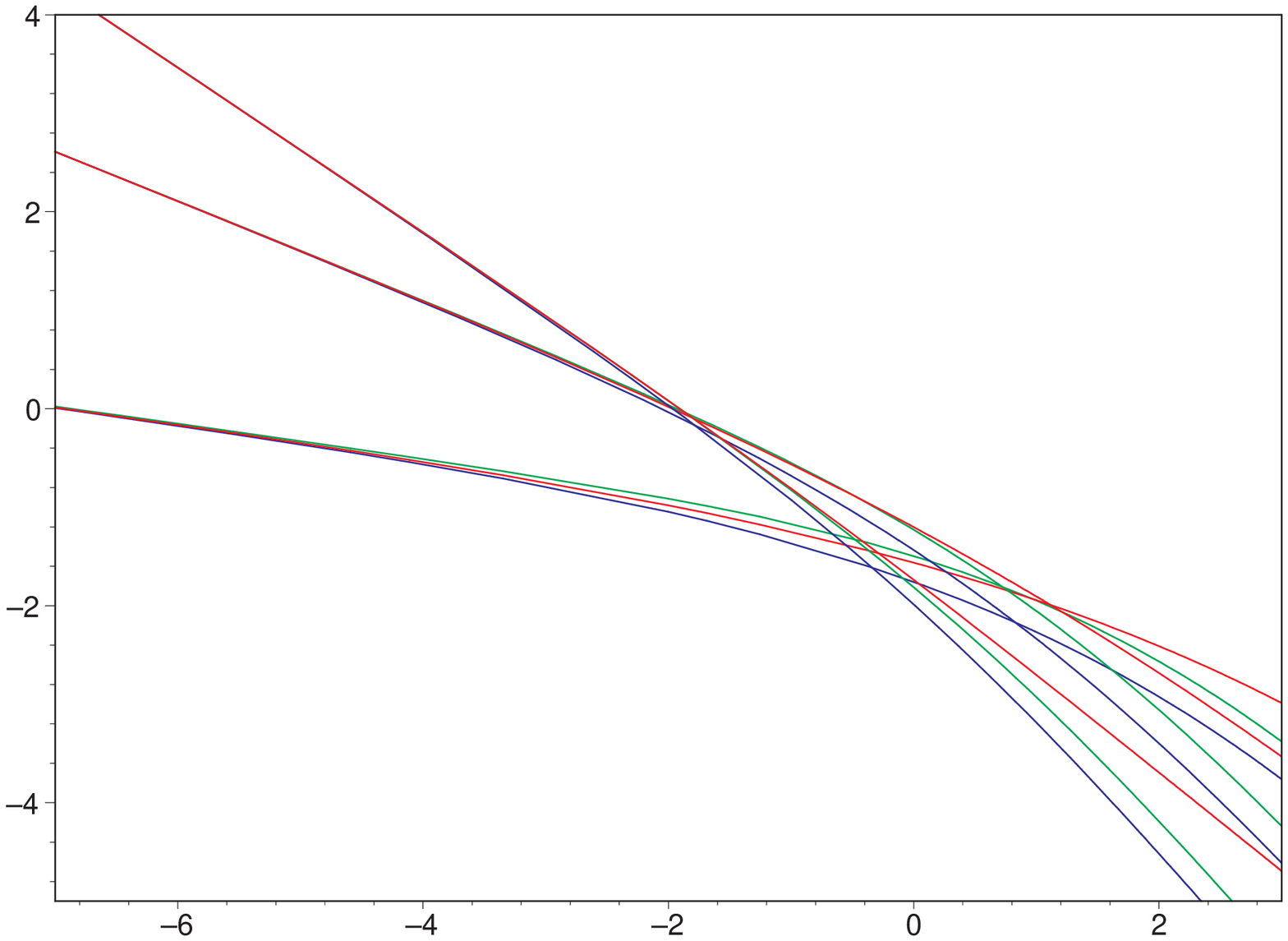,width=14cm,height=11cm}}
\put(3,6.2){{\large $g=5$}} \put(3,8.5){{\large $g=1$}}
\put(5.2,9){{\large $g=1/5$}} \put(9,0){{\large $\ln (c-c_{cr})$}}
\put(8.7,8){{\Large $\mbox{Red:}\;
\ln\left(g\sqrt{\frac{\psi(\infty)}{g+1}}\right)$}}
\put(8.7,7){{\Large $\mbox{Green:}\; \ln\left(g|D|\right)$}}
\put(8.7,6){{\Large $\mbox{Blue:}\; \ln M$}}

\end{picture}

\medskip
\caption{Dependence of the mass  and the suitably normalized
dilaton charge and $\psi(\infty)$ on $c$. When $c$ approaches the
critical value all these quantities converge to the same value
$M=g|D|=g(\psi(\infty))^{1/2}(g+1)^{-1/2}$. This corresponds to
the BPS saturation in the linear theory. } \label{bps}

\end{figure}

%%%%%%%%%%%%%%%%% N(green), xi(red) for reg
\begin{figure}
\unitlength1cm
\begin{picture}(14,11)
\put(1,0){\epsfig{file=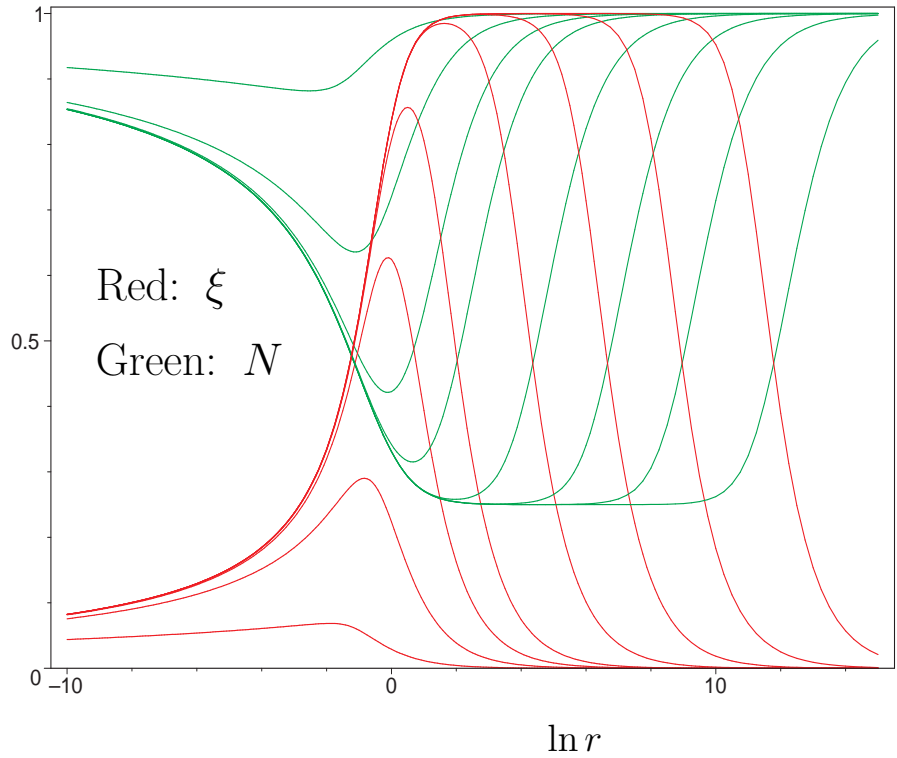,width=14cm,height=11cm}}
\put(9,0){{\large $\ln r$}} \put(3,6){{\Large $\mbox{Red:}\;\;
\xi$}} \put(3,5){{\Large $\mbox{Green:}\;\; N$}}
\end{picture}

\medskip
\caption{A family of regular solutions with $g=1$ for different
values of $c-c_{cr}$ approaching zero (to the right). One can see
that at the intermediate $r$-region both curves $N(r),\, \xi(r)$
approach the limiting (asymptotically non-flat) solution $\xi\sim
1,\, N\sim 0.25$: this region expands to the right when $c\to
c_{cr}$. Thus the effective radius of the
Einstein-Born-Infeld-dilaton particles increases when the boundary
$c=c_{cr}$ is approached. } \label{Nxir}

\end{figure}

%%%%%%%%%%%%%%%%% b_cr black holes
\begin{figure}
\unitlength1cm
\begin{picture}(14,11)
\put(1,0){\epsfig{file=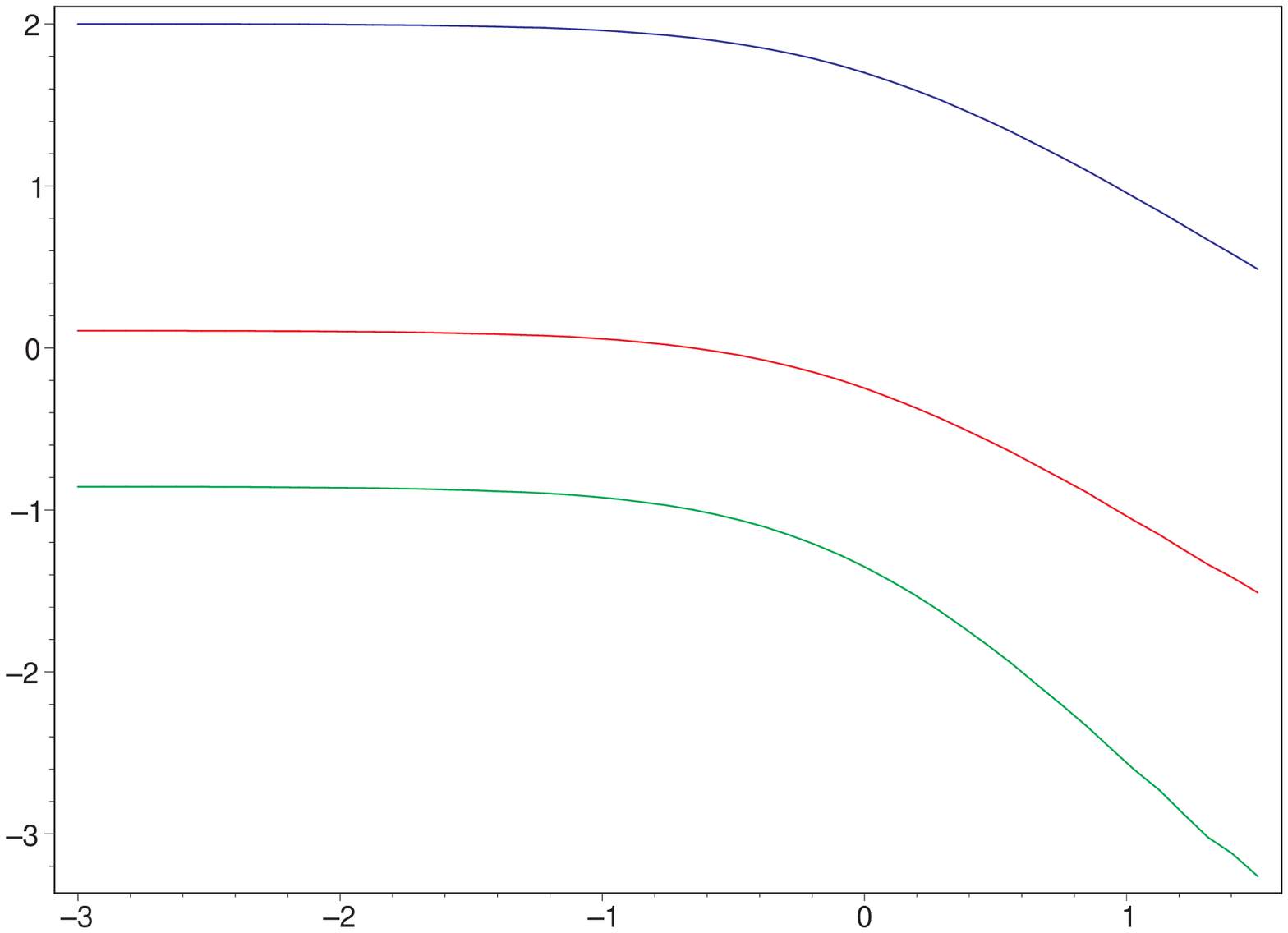,width=14cm,height=11cm}}
\put(0.4,9){{\large $\log_{10}b$}} \put(9,0){{\large $\log_{10}g$}}
\put(3,4){{\Large $r_h=0.1$}} \put(3,6){{\Large $r_h=1.0$}}
\put(3,9){{\Large $r_h=10$}}

\end{picture}

\medskip
\caption{Critical parameter $b=\psi(r_h)$ as a function of the
coupling constant for different radii of the horizon. }
\label{b_cr}

\end{figure}

%%%%%%%%%%%%%%%%% N and xi for black hole
\begin{figure}
\unitlength1cm
\begin{picture}(14,11)
\put(1,0){\epsfig{file=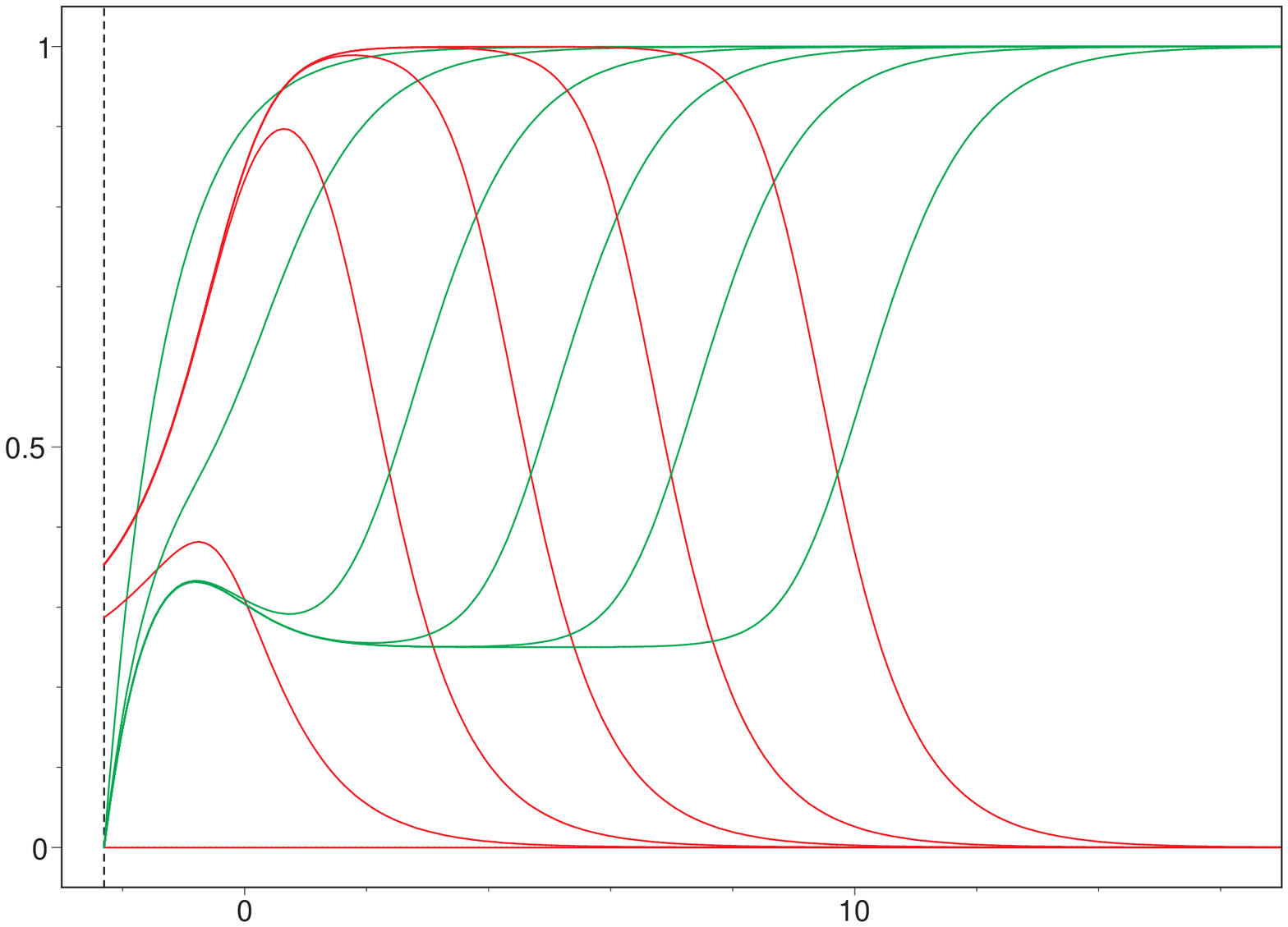,width=14cm,height=11cm}}
\put(9,0){{\large $\ln r$}} \put(3.2,2){{\large horizon}}
\put(11,6){{\Large $g=1$}} \put(11,5){{\Large $r_h=0.1$}}

\end{picture}

\medskip
\caption{A family of black hole solutions with $g=1$ for
$b_{cr}-b$ approaching zero (to the right) for 'small' black hole
$r_h=0.1$. One observes features similar to the regular case. At
the left side the behavior is modified due to the presence of the
event horizon. } \label{Nxib}

\end{figure}
%%%%%%%%%%%%%%%%% parameters delta mu nu at singularity
\begin{figure}
\unitlength1cm
\begin{picture}(14,11)
\put(1,0){\epsfig{file=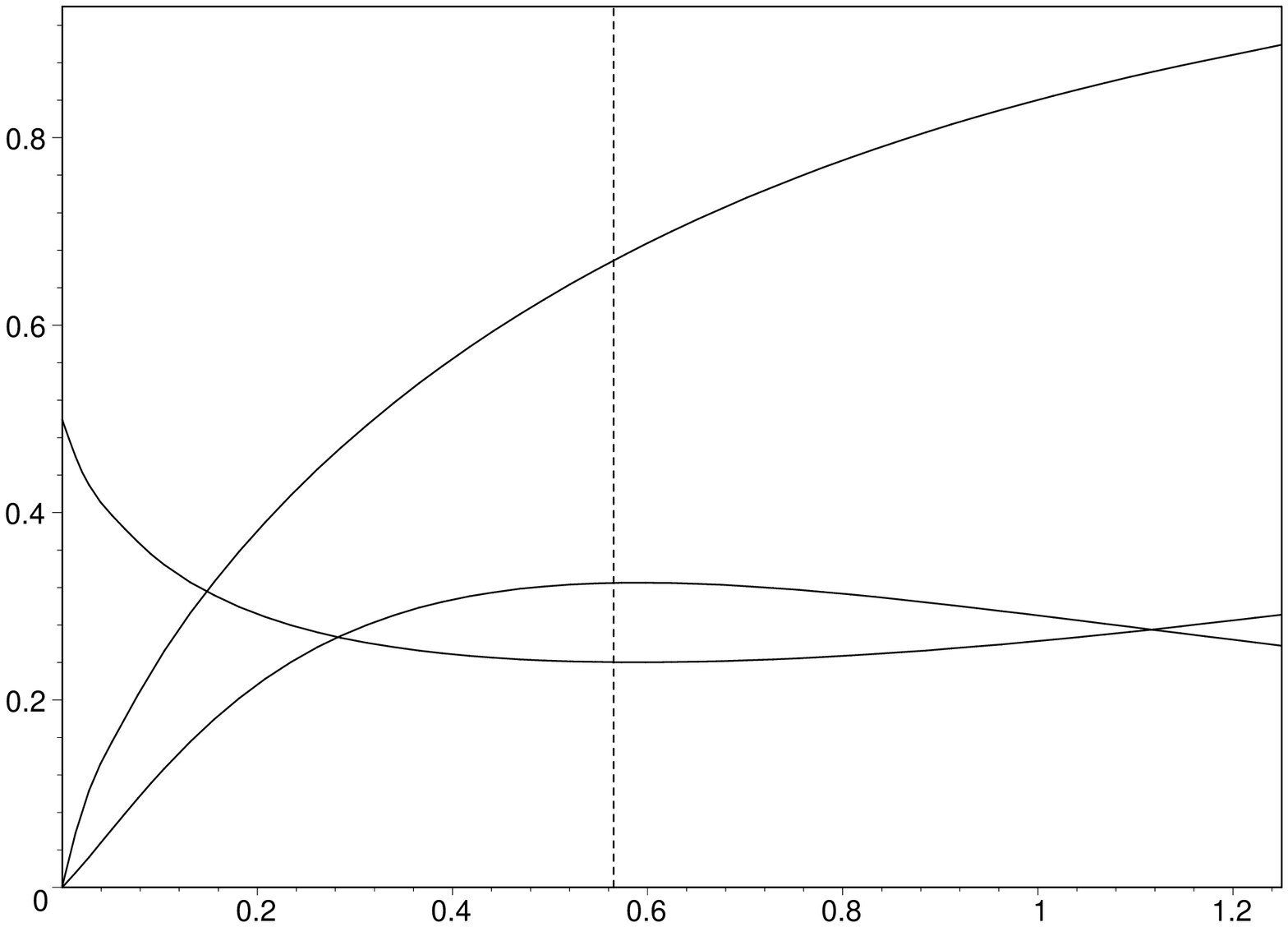,width=14cm,height=11cm}}
\put(9,0){{\large $b$}} \put(8,1.5){{\Large $b=b_{cr}$}}
\put(3,1.5){{\Large $\nu$}} \put(6,5.8){{\Large $\delta$}}
\put(3,5){{\Large $\mu$}}

\end{picture}

\medskip
\caption{Parameters of the local solution
%(\ref{es})
near the
singularity versus $b$ for $g=1,\, r_h=1$. Only the region
$b<b_{cr}$ corresponds to asymptotically flat solutions (black
holes). For $b\to 0$ one recovers the Schwarzschild solution. }
\label{sing}

\end{figure}

\end{document}